\documentclass[aps,10pt,prd,groupedaddress,preprint]{revtex4-1}
\pdfoutput=1
\usepackage{hyperref}
\usepackage{xcolor}
\usepackage{graphicx}
\usepackage{amsmath}

\begin{document}
	
	\title{Conformal Frame Dependence of Inflation}
	
	\author{Guillem Dom\`enech}
	\email[]{guillem.domenech{}@{}yukawa.kyoto-u.ac.jp}
	\author{Misao Sasaki}
	\email[]{misao{}@{}yukawa.kyoto-u.ac.jp}
	
	\affiliation{Yukawa Institute for Theoretical Physics, 
		Kyoto University, Kyoto 606-8502, Japan
	}
	
	\date{\today}
	\preprint{YITP-15-6}

\begin{abstract}
	{Physical equivalence between different conformal frames in 
scalar-tensor theory of gravity is a known fact. However, assuming that
matter minimally couples to the metric of a particular frame, which we
call the matter Jordan frame, the matter point of view of the universe
may vary from frame to frame. Thus, there is a clear
distinction between gravitational sector (curvature and scalar field) 
and matter sector. 
In this paper, focusing on a simple power-law inflation model in the
Einstein frame, two examples are considered; a super-inflationary and a 
bouncing universe Jordan frames. 
Then we consider a spectator curvaton minimally coupled to a Jordan frame,
and compute its contribution to the curvature perturbation power spectrum.
In these specific examples, we find a blue tilt at short scales for the 
super-inflationary case, and a blue tilt at large scales for the 
bouncing case.}
\end{abstract}


\maketitle

\section{Introduction \label{intro}}

Scalar fields non-minimally coupled to gravity naturally arise in higher 
dimensional theories, such us string theory (for examples 
see \cite{fujii2003scalar}), and are attractive from a renormalization 
point of view \cite{lavrov2010renormalization}. Such effective field theory 
can be described within the framework of Scalar-tensor theory of gravity, 
first introduced by Jordan \cite{jordan1959gegenwartigen} and followed by 
Brans and Dicke \cite{brans1961mach}, who realised that by means of 
a field dependent conformal transformation, i.e. a field dependent re-scaling
of the metric, the non-minimal coupling can be absorbed and we are left with 
the usual Einstein-Hilbert action with a scalar field.
Thus, this led to the notion of two distinct frames; the Einstein frame 
where the scalar is minimally coupled, and a Jordan frame where
a non-minimal coupling is present. Here and throughout the paper, 
we define the Jordan frame as the one in which matter fields 
are minimally coupled with the metric of the frame. Conversely, matter
fields have a universal dilatonic coupling with the scalar field in the
Einstein frame.

There have been many controversial arguments about the physical 
equivalence of conformal frames and much effort has been made to 
clarify the situation \cite{makino1991density, faraoni2007pseudo, deruelle2011conformal, gong2011conformal, white2012curvature, jarv2014invariant,catena2007einstein,chiba2013conformal,chiba2008extended,
qiu2012reconstruction,li2014generating}. The general conclusion at the classical level is that although physically equivalent, interpretations differ from frame to frame. On the contrary, 
it is still unclear whether they are equivalent at the quantum 
level \cite{george2014quantum} or not \cite{kamenshchik2014frame}. 
To our opinion it seems it needs quantum gravity to settle down this
problem at the quantum level.

Inflation is widely accepted as the model of the early universe and is
supported by current data \cite{hinshaw2013nine,ade2013planck,ade2013planck2}. 
As scalar fields play an important role in driving inflation, 
it is of interest to consider the consequences and implications of
the case when the inflaton field is non-minimally coupled to gravity. 
More specifically, we are interested in the case where there is a Jordan 
frame in which the inflaton field is non-minimally coupled while
matter fields are minimally coupled.

However, since there are functional degrees of freedom in the form
of the non-minimal coupling, it does not give us any useful insight
if we stick to the general case. On the other hand, if we restrict
our consideration too strongly, then we would not be able to learn much 
from it. Here, we focus on a simple but exact model which can be
treated analytically, yet allows sufficient varieties in its outcome.
Namely, we choose the power-law inflation model~\cite{lucchin1985power}
which is realised by an exponential potential. We assume this is what
we have in the Einstein frame. In this case, the general solution is
known analytically~\cite{russo2004exact,andrianov2011general}.

The purpose of this work is to study how the physics from the matter 
point of view may vary from frame to frame in the explicit case of 
power-law inflation. For this purpose a simple way is to start in the 
Einstein frame and, by means of a conformal transformation, we go to
a Jordan frame where matter is defined to be minimally coupled to
the metric.

The structure of this work is as follows. 
In section \ref{sec:pl} we briefly review the general
conformal transformation in a scalar-tensor theory,
and apply it to power-law inflation.
In section \ref{sec:matter} we introduce a curvaton to our
model, and study its behaviour in a Jordan frame. It turns out 
that this simple curvaton model can give rise to an interesting 
physics from the matter point of view, and therefore may generate
interesting features in the CMB angular spectrum.
Finally in section \ref{sec:conclusion} we summarise the result
and discuss its possible imprints in observational data.

\section{Power-law inflation} 
\label{sec:pl}

The action for a tensor-scalar theory in the Einstein frame reads
\begin{eqnarray}
&&S_g=
\int d^4x \sqrt{-g}\left(\frac{M_{pl}^2}{2}R
-\frac{1}{2}g^{\mu\nu}\partial_\mu\varphi\partial_\nu\varphi
-V[\varphi]\right),\label{eh}
\end{eqnarray}
where $M_{pl}$ is the Planck mass, $\varphi$ is an inflaton field and the sub-index $S_g$ stands for the gravitational sector. 
After an arbitrary conformal transformation, 
\begin{eqnarray}
g_{\mu\nu}= F[\varphi]\tilde{g}_{\mu\nu}, \label{conftrans}
\end{eqnarray}
where $F[\varphi]$ is a well-behaved non-zero function, we obtain the corresponding action in a Jordan frame,
\begin{eqnarray}
&&S_g=\int d^4x \sqrt{-\tilde{g}} 
\left(\frac{M_{pl}^2}{2}F[\tilde{\varphi}]\tilde{R}
-\frac{1}{2}\tilde{g}^{\mu\nu}\partial_\mu\tilde{\varphi}
\partial_\nu\tilde{\varphi}-\tilde{V}[\tilde{\varphi}]\right),
\label{jaction}
\end{eqnarray}
where the scalar field has been redefined by
\begin{eqnarray}
\left(\frac{d\tilde{\varphi}}{d\varphi}\right)^2=F[\varphi]
\left|1-\frac{3}{2}M_{pl}^2\left(\frac{\partial \ln F}
{\partial\varphi}\right)^2\right|
\label{field}
\end{eqnarray}
and the new potential is
\begin{eqnarray}
\tilde{V}[\tilde{\varphi}]
=V[\varphi(\tilde{\varphi})]F^{2}[\varphi(\tilde{\varphi})].
\end{eqnarray}
According to \cite{gong2011conformal,white2012curvature} both actions 
lead to the same curvature power spectrum and, hence, they are 
indistinguishable observationally, for example in the Cosmic Microwave 
Background. Besides, regarding the running of the 
units \cite{dicke1962mach,faraoni2007pseudo} we assume that after inflation
the inflaton settles down to its minimum. As a result, the Einstein and 
Jordan frames become equivalent.

Thus, one may wonder why should we consider Jordan frames if the Einstein 
frame is much simpler. In fact, once we take matter into account,
one can usually define a frame where matter is minimally 
coupled to the metric \eqref{mpv} and therefore, as matter is concerned, physical interpretations in that frame are straightforward. 
In this sense, the gravitational sector, namely the terms composed of
the scalar curvature and the inflaton, is generally physically independent 
of the \textit{arbitrary} re-scaling function $F[\varphi]$, while 
the matter sector is not. 

To illustrate this, we consider the Jordan frame where a matter field 
$\chi$ is minimally coupled to the metric,
\begin{eqnarray}
S_m=\int d^4x \sqrt{-\tilde{g}}{\cal{L}}_m[\tilde{g}^{\mu\nu},\chi].
\label{mpv}
\end{eqnarray}
Transforming back to the Einstein frame \eqref{conftrans},
we are left with a non-minimal coupling of the matter with the 
inflaton through $F[\varphi]$,
\begin{eqnarray}
S_m=\int d^4x \sqrt{-g}F^{-2}{\cal{L}}_m[F[\varphi]g^{\mu\nu},\chi].
\end{eqnarray}

Power-law inflation was first introduced by Lucchin 
and Matarrese ~\cite{lucchin1985power} where it was shown that a scalar field 
with an exponential potential,
\begin{eqnarray}
V[\varphi]=V_0e^{-\lambda\varphi/M_{pl}},
\end{eqnarray}
in a flat FLRW background,
\begin{eqnarray}
ds^2=-dt^2+a^2d\textbf{x}^2,
\end{eqnarray}
give rise to an exact power-law solution,
\begin{eqnarray}
a=a_0(t/t_0)^p\quad (0<t<\infty)\,,
\end{eqnarray}
where $p=2/\lambda^2$ and
 $\lambda^2V_0t^2_0=2M^2_{pl}(3p-1)$.
As a result, this solution describes an initial big bang followed 
by an eternal expansion. The Hubble and slow-roll parameters are
given respectively by
\begin{eqnarray}
H\equiv\dot{a}/a=p/t\,,\label{hubble}
\\
\epsilon\equiv -\dot H/H^2=1/p\,,
\end{eqnarray}
where a dot refers to the derivative with respect to the proper time $t$. 
The solution to the scalar field equation,
\begin{eqnarray}
\ddot{\varphi}+3H\dot{\varphi}+V,_{\varphi}=0,
\end{eqnarray}
is given by
\begin{eqnarray}
\varphi=\frac{2M_{pl}}{\lambda}\ln(t/t_0)\,,
\label{phi}
\end{eqnarray}
Furthermore, comparing \eqref{hubble} and \eqref{phi} we are led to 
the equality, 
\begin{eqnarray}
\lambda H M_{pl}=\dot{\varphi}\label{lhphi},
\end{eqnarray}
which will be used later.

\begin{figure}[tbp]
\centering
\includegraphics[width=0.6\columnwidth]{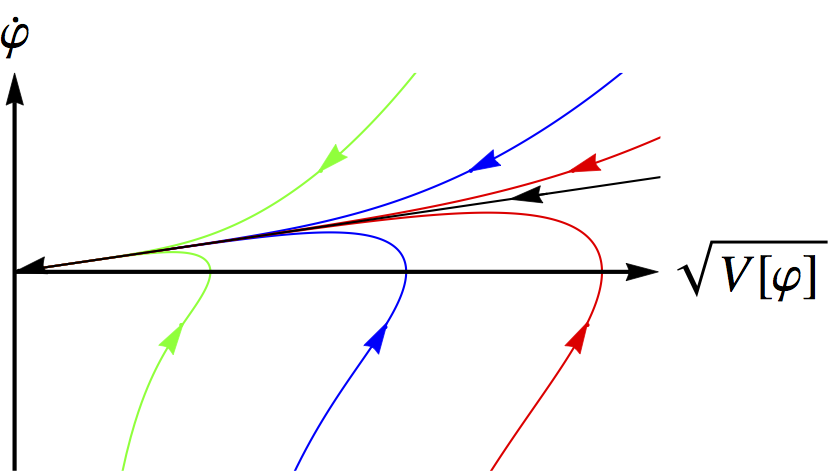}
\caption{Phase space diagram for a particular $\lambda$. Black line 
stands for the attractor power-law solution, whereas the colored lines 
are the two families of general solutions (rolling-up solutions
below the black line and rolling-down solutions above the black 
line).}
\label{diagram}
\end{figure}

Following the work of \cite{russo2004exact,andrianov2011general}, 
one can check that $p>1$ corresponds to an inflationary attractor solution,
as it can be seen from Fig.~\ref{diagram}. Likewise, one can check that 
there are essentially two types of general solutions corresponding to two 
distinct initial conditions, i.e. the field starts rolling up or the field 
starts rolling down the potential. We decided not to consider these
solutions further in this work due to not-well-defined initial conditions 
when computing the curvature perturbation power spectrum. 
For a detailed review of the solutions,
see \cite{russo2004exact,andrianov2011general}.

\subsection{Primordial fluctuations}

The next stage is to compute the primordial power spectrum for 
this attractor solution, with the help of cosmological
perturbation theory~\cite{kodama1984cosmological,mukhanov1992theory}.
First, we consider the Mukhanov-Sasaki equation for the curvature 
perturbation~\cite{Mukhanov:1985rz,sasaki1986large},
\begin{eqnarray}
r''_k+2\frac{z'}{z}r'_k+k^2r_k=0,
\end{eqnarray}
where a prime denotes derivative with respect to the conformal
 time $d\eta=dt/a$ and
\begin{eqnarray}
z=a\varphi'/{\cal H}=\lambda a M_{pl},
\end{eqnarray}
where ${\cal H}=a'/a$ and we used \eqref{lhphi} for the last equality. 
Second, we assume that the field is in slow-roll regime, i.e. $p\gg1$, 
and use the WKB approximation at $k\gg{\cal H}$,
yielding the known results \cite{sasaki1986large},
\begin{eqnarray}
{\cal P}_{{\cal{R}}_c}(k)=\frac{4\pi k^3}{(2\pi)^2}|r_k|^2
=\left(\frac{H^2}{2\pi\dot{\varphi}}\right)^2
=\frac{p}{8\pi^2}\frac{H^2}{M^2_{pl}}
\end{eqnarray}
at $k={\cal H}=Ha$, namely at the horizon crossing time.
Rewriting the above in terms of $k$ gives the primordial 
curvature perturbation spectrum,
\begin{eqnarray}
{\cal P}_{{\cal{R}}_c}(k)=\frac{p}{8\pi^2}\frac{H_0^2}{M^2_{pl}}
\left(\frac{k}{k_0}\right)^{\frac{-2}{p-1}}
\label{scalar},
\end{eqnarray}
where $k_0$ is a reference wavenumber that crosses the horizon at
$t=t_0$, and $H_0=p/t_0$.
Similarly, the tensor perturbation spectrum is given by
\begin{eqnarray}
{\cal P}_{{\cal{T}}}(k)=\frac{4\pi k^3}{(2\pi)^2}|h_k|^2
=\frac{2}{\pi^2}\frac{H^2}{M^2_{pl}}
=\frac{16}{p}{\cal P}_{{\cal{R}}_c}(k),
\end{eqnarray}
in agreement with the results of standard power-law 
inflation \cite{lucchin1985power}. 
For a more precise result, see \cite{lyth1992curvature}.

Here our main motivation to consider power-law inflation is
because the model allows us to study it analytically.
Nevertheless, it may be of some interest to check
the current status of the observational constraints.
The spectral index is given in terms of the parameter $p$ as
\begin{eqnarray}
n_s-1=\frac{d\ln{\cal P}_{{\cal{R}}_c}}{d\ln k}=\frac{-2}{p-1}\,,
\end{eqnarray}
which is slightly red, while the tensor to scalar ratio is given by
\begin{eqnarray}
r=\frac{{\cal P}_{{\cal{T}}}}{{\cal P}_{{\cal{R}}_c}}=\frac{16}{p}\,.
\label{r}
\end{eqnarray}
The most recent observational constraints are by Planck, 
which gives $n_{s,planck}\approx 0.96$ and $r_{planck}<0.1$. 
 From the former we obtain $p\approx 50$, but this gives
$r\approx 0.32$ which is too large. 
This is a rather common feature of large field inflation models.
However, this discrepancy can be alleviated by invoking a 
curvaton~\cite{moroi2002cosmic,enqvist2002adiabatic,lyth2002generating}.
Although our purpose is not to resolve the discrepancy but to study 
the matter's physics in the Jordan frame, regarding
a curvaton as a representative of matter, it turns out that we can
actually make power-law inflation observationally more attractive,
as will be shown below.

\section{The matter point of view}
\label{sec:matter}

In order to understand the matter point of view, we consider an almost massless curvaton $\chi$~\cite{moroi2002cosmic,enqvist2002adiabatic,lyth2002generating} with sub-dominant energy density
which plays no role in driving inflation. Therefore 
the inflationary dynamics is dictated by the inflaton $\varphi$. 
The action of the spectator curvaton takes the usual form,
\begin{eqnarray}
S_m=\int d^4x \sqrt{-\tilde{g}}
\left(-\frac{1}{2}\tilde{g}^{\mu\nu}
\partial_\mu\chi\partial_\nu\chi-\frac{1}{2}\tilde{m}^2\chi^2\right)
\end{eqnarray}
where $\tilde{g}_{\mu\nu}$ is the metric of a particular Jordan
frame and $\tilde{m}$ is the mass of the curvaton in that frame. 
In general, after a conformal transformation \eqref{conftrans} 
the background is modified,
\begin{eqnarray}
d\tilde{s}^2=F^{-1}[\varphi]ds^2=-d\tilde{t}^2
+\tilde{a}^2d\textbf{x}^2,
\end{eqnarray}
and the proper time and the scale factor are respectively redefined as
\begin{eqnarray}
d\tilde{t}=F^{-1/2}dt\label{jtime}
\end{eqnarray}
and
\begin{eqnarray}
\tilde{a}=F^{-1/2}a.\label{ja}
\end{eqnarray}
Consequently, the new conformal Hubble parameter is
\begin{eqnarray}
\tilde{\cal H}={\cal H}-F'/2F={\cal H}
\left(1-\frac{\lambda}{2}\frac{\partial\ln F}{\partial \varphi/M_{pl}}\right),
\label{jhubble}
\end{eqnarray}
where \eqref{lhphi} has been used. 

In passing, for this simple case, it may be worth noting how the 
frame independence of the quantization is realized inspite of the 
difference in the physical interpretation. 
First of all, if the conformal time is used as the time coordinate,
the invariance of the canonical commutation relation is trivial. 
Consequently the invariance of the field equation is also trivial.
Nevertheless it is instructive take a look at the
differential equation for the mode function in each frame.
In the Jordan frame it is
\begin{eqnarray}
\chi''_k+2\tilde{\cal H}\chi'_k+(k^2+\tilde{m}^2\tilde{a}^2)\chi_k=0.
\label{jmode}
\end{eqnarray}
In the Einstein frame where the action is
\begin{eqnarray}
S_m=\int d^4x \frac{\sqrt{-g}}{F}\left(-\frac{1}{2}g^{\mu\nu}
\partial_\mu\chi\partial_\nu\chi-\frac{1}{2}m^2\chi^2\right),
\end{eqnarray}
where we redefined the mass by $m=F^{-1/2}\tilde{m}$, the equation for the 
mode functions reads
\begin{eqnarray}
\chi''_k+2({\cal H}-F'/2F)\chi'_k+(k^2+m^2a^2)\chi_k=0.
\label{emode}
\end{eqnarray}
Comparing both \eqref{jmode} and \eqref{emode}, it is clear that they are 
exactly the same, but their interpretations are rather different. 
In fact, while \eqref{jmode} is the usual differential equation for the 
mode functions of a canonical scalar field in a FLRW background,
\eqref{emode} contains explicitly the effect of the non-canonical coupling 
with the inflaton in the kinetic term.

In any case, the resulting curvaton perturbation spectrum is frame independent.
Keeping this in mind, we consider a couple of particular examples below
separately. To begin with, we assume a slow-roll inflationary Einstein frame, 
i.e. $p\gg 1$, unless otherwise noted.

\subsection{Power-law Jordan frame\label{case1}}

\begin{figure}[tbp]
\centering
\includegraphics[width=0.6\columnwidth]{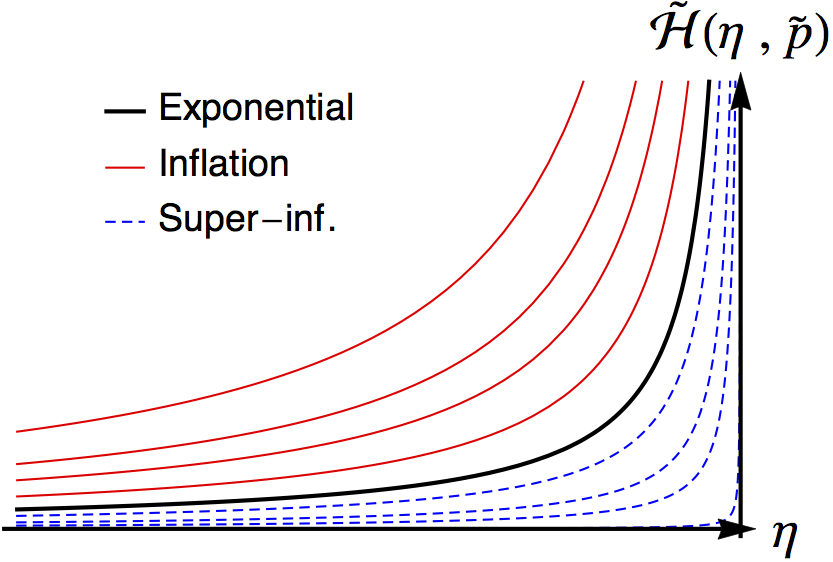}
\caption{Jordan conformal Hubble parameter as a function of conformal 
time $\eta$. The black line is the exponential expansion 
$\tilde{p}\rightarrow\infty$, orange lines stand for power-law inflation, 
$\tilde{p}>1$, whereas green dashed lines stand for super-power-law 
inflation, $\tilde{p}<0$.}
\label{jhub}
\end{figure}

First we consider a simple conformal transformation,
\begin{eqnarray}
F(\varphi)=\left[V(\varphi)\right]^{-\gamma}
=\textrm{e}^{\gamma\lambda\varphi/M_{pl}}=(t/t_0)^{2\gamma},\label{fcase1}
\end{eqnarray}
inspired by dilaton models in string theory, for example \citep{blumenhagen2007four}. 
After integrating \eqref{field} and substituting it into the 
Jordan action \eqref{jaction}, we are led to
\begin{align}
S=\int d^4x \sqrt{-\tilde{g}} 
\left(\frac{\xi\tilde{\varphi}^2\tilde{R}}{2M^4_{pl}}
-\frac{1}{2}\tilde{g}^{\mu\nu}
\partial_\mu\tilde{\varphi}\partial_\nu\tilde{\varphi}
-V_0\Big(\frac{\xi\tilde{\varphi}^2}{M_{pl}^2}\Big)^{2-\frac{1}{\gamma}}\right),
\end{align}
where 
\begin{align}
\xi=\frac{(\gamma\lambda)^2}{4-6(\gamma\lambda)^2}
=\frac{\gamma^2}{2(p-3\gamma^2)}\,.
\label{xivalue}
\end{align}
Here we note that for $\gamma^2>2/(3\lambda^2)=p/3$, the gravitational
part of the Jordan frame action becomes ghost-like. Nevertheless 
since the original Einstein frame action is perfectly normal,
the system is perfectly stable in spite of its seemingly disastrous
appearance~\cite{Maeda:1988ab}.

In this Jordan frame, 
we encounter another power-law with a different power law index
\begin{eqnarray}
\tilde{a}=a_0(\tilde{t}/\tilde{t}_0)^{\tilde{p}},\label{ja2}
\end{eqnarray}
in agreement with \cite{li2014generating}, where
the condition to obtain the scale-invariant tensor spectrum was 
discussed from the Jordan frame point of view.
In \eqref{ja2} we have integrated Jordan time $\tilde{t}$ \eqref{jtime}
 and replaced into the new scale factor \eqref{ja}, where the 
Jordan time $\tilde{t}$ now runs from $0$ to $\infty$ for 
$\tilde{p}>1$ ($\gamma<1$) and from $-\infty$ to $0$ for
$\tilde{p}<1$ ($\gamma>1$).
Correspondingly, the Jordan power-law index and $\tilde{t}_0$ are 
respectively related to those in the Einstein frame by
\begin{eqnarray}
\tilde{p}-1=\frac{p-1}{1-\gamma}\label{ptilde}
\end{eqnarray}
and
\begin{eqnarray}
\tilde{t}_0=\frac{t_0}{1-\gamma}.\label{t0}
\end{eqnarray}
It is interesting to note that from \eqref{ptilde} this Jordan frame 
is not restricted to $\tilde{p}>1$ nor even $\tilde{p}>0$. Thus, 
although an almost scale invariant spectrum is obtained independent
of the frame, this Jordan frame may not be seen as an inflationary universe from the curvaton point 
of view, which is subject to the Jordan frame metric. This result reminds 
us of \cite{tsujikawa2000power}, where the non-minimal coupling is used to ``assist'' inflation. In other words inflation is recovered in the
Einstein frame even though the Jordan frame is not inflationary,
thanks to the non-minimal coupling of the inflaton.

In this Jordan power-law, we encounter three general cases.
First, for $\gamma<1$, we have $\tilde{p}>1$ and the curvaton also feels 
inflation with a different power-law index. The case $\gamma=1$ corresponds to
the exact exponential expansion.
Second, for $1<\gamma<p$ we have $\tilde{p}<0$ and the curvaton experiences
a super-inflationary universe, where by super-inflationary we
mean that the universe expands faster than an exponential expansion.
The behavior of the conformal Hubble parameter for $0<\gamma<p$
($\tilde{p}<0$ and $\tilde{p}>1$) is illustrated in Fig.~\ref{jhub}.
Finally, for $\gamma>p$, we have $0<\tilde{p}<1$ and
the curvaton is in a decelerated contracting universe.
Note that $\gamma>p$ implies $\gamma^2>p/3$ for $p>1/3$.
Hence this last case corresponds to the ghost-like gravity mentioned
before.

\begin{figure}[tbp]
\centering
\includegraphics[width=0.6\columnwidth]{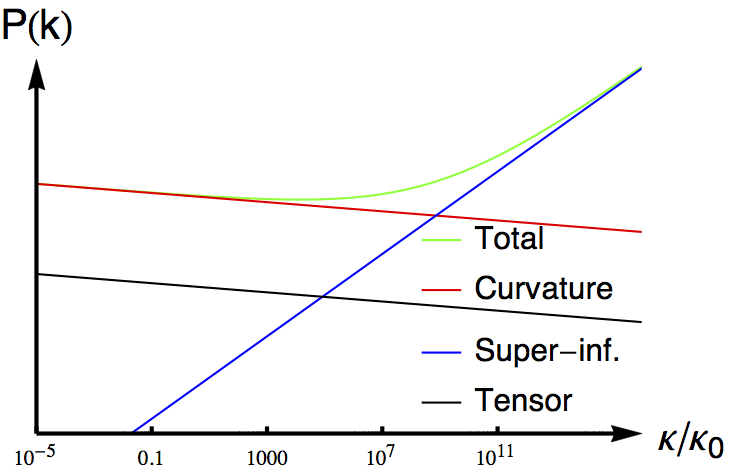}
\caption{Power spectrum in logarithmic scale for the power-law Jordan case. Inflaton 
curvature power spectrum (red line), tensor power spectrum (black line),
 super-inflationary curvaton power spectrum (blue line) and total 
scalar power spectrum (green line).}
\label{blue}
\end{figure}

In order to compute the scalar power spectrum in this case we assume 
again that we are in the slow-roll regime, i.e. $|\tilde{p}|\gg1$, 
and use the WKB approximation inside the horizon and
assume that a mode freezes out instantaneously at horizon crossing,
which we call the instantaneous horizon exit assumption. 
The curvature perturbation spectrum due to the curvaton, under the sudden curvaton decay approximation \citep{lyth2002generating}, is given by
\begin{eqnarray}
{\cal P}_{\chi}(k)=r_\star\frac{\delta\chi^2}{\chi_\star^2}=r_\star\frac{4\pi k^3}{(2\pi)^2}\frac{|\chi_k|^2}{\chi_\star^2}, \label{powerchi}
\end{eqnarray}
where $\chi_\star$ is the background value of the curvaton field and $r_\star=\rho_{\chi_\star}/(\rho+p)_{{\rm tot}_\star}$ is the energy density fraction of the curvaton at the time of decay. Note that for \eqref{powerchi} to be valid the curvaton must have a non-vanishing background value, which in turn implies $\tilde{H}\gg \tilde{m}$ \citep{lyth2002generating,bartolo2002simplest}. This condition is readily seen from the field equations of motions, i.e.
\begin{align}
\ddot{\chi}+3\tilde{H}\dot{\chi}+\tilde{m}^2\chi^2=0,
\end{align}
where requiring $\tilde{H}\gg \tilde{m}$ prevents the curvaton to settle down to its minimum.
With the above assumptions we recover
the usual scalar power spectrum,
\begin{eqnarray}
{\cal P}_{\chi}(k)=r_\star\frac{\tilde{H}^2}{(2\pi M_{pl}\chi_\star)^2}
=r_\star\frac{\tilde{H}_0^2}{M^2_{pl}\chi_\star^2}
\left(\frac{k}{k_0}\right)^{\frac{-2}{\tilde{p}-1}}\,,
\label{powerchi2}
\end{eqnarray}
where the total scalar power spectrum is expressed as
\begin{eqnarray}
{\cal P}_{\rm tot}={\cal P}_{{\cal{R}}_c}+{\cal P}_{\chi}.
\end{eqnarray}
%where $\chi_\star$ is determined by the energy density of the curvaton
%when it decays. 
As a result, the new tensor to scalar ratio \eqref{r}
 is given by \cite{fujita2014curvaton} 
\begin{eqnarray}
\tilde{r}=\frac{r}{1+{\cal P}_{\chi}/{\cal P}_{{\cal{R}}_c}}\,.
\label{rtilde}
\end{eqnarray}
If we assume the curvaton energy density when it decays
to be comparable to or greater than that due to the inflaton,
$\tilde{r}$ becomes small enough and the non-gaussianity parameter, $f_{NL}$, 
becomes $O(1)$ or smaller \cite{enqvist2013mixed, byrnes2014comprehensive},
making this scenario more consistent with the current observational data.

The curvaton spectral index can be easily extracted from \eqref{powerchi2}.
We obtain
\begin{eqnarray}
\tilde{n}_\chi-1=\frac{-2}{\tilde{p}-1}\,,
\end{eqnarray}
which need not be a red index. 
In particular, a blue tilt can be naturally achieved for $\tilde{p}<0$, 
that is the super-inflationary situation, which is a common feature of super-inflationary models \cite{piao2004inflation,piao2004phantom,cicoli2014just,biswas2014super,liu2013cmb,piao2003nearly}. See Fig.~\ref{blue}.
This blue spectrum contribution can be important on small scales.
For example, it may enhance the primordial black hole formation which can
account for a fair amount of dark matter \cite{green2014primordial}. 
One may fairly wonder how important the back-reaction due to the 
large curvaton fluctuations would be. This would become important
when the amplitude of the power spectrum became of order unity.
However, before the back-reaction would become important, one
would probably encounter an over-abundance of primordial black holes.
This means that there should be a cutoff. Study on such a case
may be of interest but it is beyond the scope of the present paper.

\subsection{Jordan bouncing universe\label{case2}}

\begin{figure}[tbp]
\centering
\includegraphics[width=0.6\columnwidth]{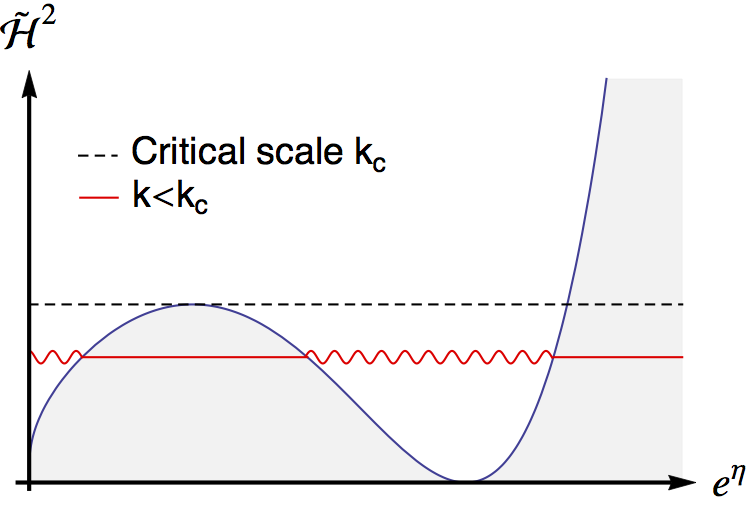}
\caption{Conformal Jordan squared hubble parameter (blue line) for the
 bouncing universe case and horizontal lines are constant $k$ lines. 
Shadowed region stands for super-horizon regime and the consequent 
freezing of the modes.}
\label{bounce}
\end{figure}

As a second example, we consider a conformal transformation of the type,
\begin{eqnarray}
F(\varphi)=\Big(1+\textrm{e}^{\frac{-\gamma\lambda}{2M_{pl}}\varphi}\Big)^{-2}
=\Big(1+(t/t_0)^{-\gamma}\Big)^{-2}\,.
\label{fcase2}
\end{eqnarray}
For $\gamma>0$ this reproduces the first example, subsection \ref{case1}, 
at early times ($t\ll t_0$) and is just identical to the Einstein frame 
at late times ($t\gg t_0$), and it is the other way around for $\gamma<0$. 
From now on, we focus our attention on the case $\gamma>1$ where 
the approximate behaviour of the scale factor is
\begin{align}
\tilde{a} \approx \left\{
\begin{array}{ccc}
a_0(-\tilde{t}/\tilde{t}_0)^{\tilde{p}} & \quad |\tilde{t}|\gg\tilde{t_0}
& \quad(\tilde{t}<0)
\\
a_0(\tilde{t}/\tilde{t}_0)^{p} & \quad \tilde{t}\gg\tilde{t_0}
\end{array}
\right.,
\end{align}
where the Jordan time, $\tilde{t}$ \eqref{jtime}, runs from $-\infty$ to 
$\infty$. 

Most interestingly, for the case where $0<\tilde{p}<1$ the 
curvaton is in a bouncing universe with bounce at
\begin{eqnarray}
t_{bounce}=t_0\left(\frac{\gamma}{p}-1\right)^{1/\gamma}.
\label{tbounce}
\end{eqnarray}
Note that the big bang singularity in the Einstein frame is sent to
$\tilde{t}\to-\infty$ while the bounce occurs at a perfectly regular epoch
in the Einstein frame.
This model differs from usual bouncing cosmologies       \cite{allen2004cosmological,lyth2002primordial,cai2007bouncing,qiu2011bouncing} 
(for a review of bouncing cosmologies see \cite{battefeld2014critical}) 
in that the initial singularity is avoided from the matter point of view 
but it is still present in the gravitational sector. 
In order not to induce any confusion, let us call this a Jordan bouncing
universe. In essence, the inflaton is preventing matter to feel the 
initial singularity through its non-canonical coupling. 
As a result, in this simple model the curvaton is in a bouncing
 universe where one can compute 
the resulting power spectrum due to a well defined initial vacuum state 
as $\tilde{\cal H}$ vanishes in the limit $\eta\rightarrow -\infty$,
\begin{eqnarray}
\tilde{\cal H}={\cal H}\left(1-\frac{\gamma}{p}
\frac{\textrm{e}^{\frac{-\gamma\lambda}{2}\varphi}}
{1+\textrm{e}^{\frac{-\gamma\lambda}{2}\varphi}}\right).
\end{eqnarray}

Before ending this section, we compute the curvaton power spectrum for 
the Jordan bouncing universe. It should be noted that, due to the bounce 
at $\tilde{\cal H}=0$, and a change from a decelerated contracting phase 
to an accelerated expanding phase at $\tilde{\cal H}'=0$, 
there appears a new critical scale $k_c$. See Fig.~\ref{bounce}. 
As schematically shown in the figure, the large scale modes $k<k_c$ go 
out of the Hubble horizon, re-enter the horizon before bounce, 
and go out of the horizon again during the final inflationary stage, 
while the small scale modes $k>k_c$ remain inside the horizon all through
the contractig and bouncing stages until the final inflationary stage.

Moreover, in order for the curvaton to contribute to the scalar 
power spectrum, it must have a non-vanishing background $\chi=\chi_0(\eta)$,
which implies that $\tilde{H}\gg \tilde{m}$ has to be satisfied not only
at late times but also at early times. We may achieve this 
condition by assuming an inflaton dependence in the mass of the curvaton, 
at least $\tilde{m}[\varphi]\propto F^{1/2}[\varphi]$ at early times.
\begin{figure}[tbp]
\centering
\includegraphics[width=0.7\columnwidth]{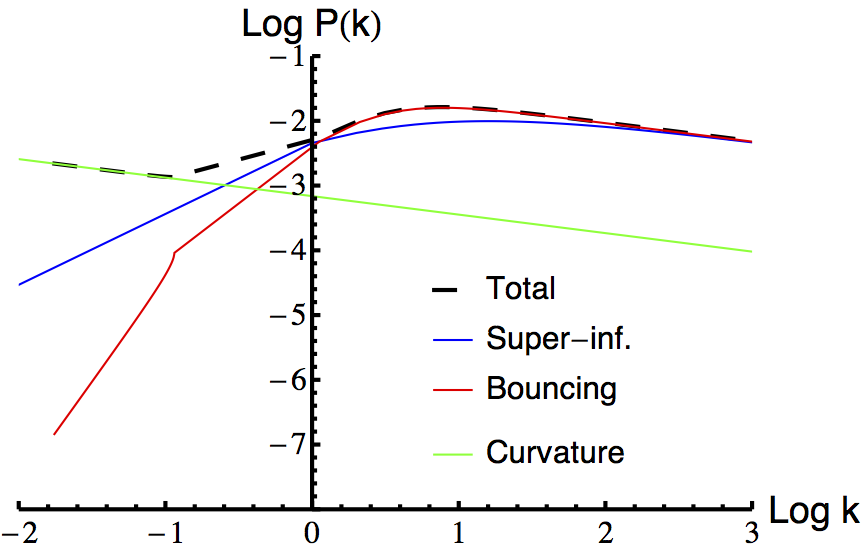}
\caption{Power spectrum for the bouncing universe (red line), 
super-inflationary to inflationary universe (blue line), curvature power spectrum (green line) and total scalar power spectrum for the bouncing case (black line).}
\label{powerfinal}
\end{figure}
In this way, with the instantaneous horizon exit and reentry 
approximation for $k<k_c$, \footnote{It is not a good approximation 
for modes near $k_c$ but the general behaviour is not changed.} 
the mode functions for the curvaton when they are inside the horizon
at the final inflationary stage are given by
\begin{align}
\chi_k \approx \left\{
\begin{array}{cc}
\frac{1}{\sqrt{2k}}\frac{\tilde{a}(\eta_2)}{\tilde{a}(\eta_1)}
\frac{1}{\tilde{a}(\eta)} \textrm{e}^{-\textrm{i}k\eta+i\alpha_k}
&\quad k<k_c\\
\frac{1}{\sqrt{2k}}\frac{1}{\tilde{a}(\eta)} \textrm{e}^{-\textrm{i}k\eta}
& \quad k>k_c
\end{array}
\right.,
\end{align}
where $\eta_1$ and $\eta_2$ are, respectively, the horizon exit and 
reentry conformal times for modes with $k<k_c$. Although the instantaneous
 horizon exit and reentry approximation would break down for $\tilde{p}\ll1$,
the resulting power spectrum can be still considered as a rough 
approximation.
Again, making use of \eqref{powerchi} we obtain the power spectrum
which we compute numerically because of the non trivial form of the 
scale factor around the bounce. The result is presented in Fig.~\ref{powerfinal}.
As may be naively expected, the spectrum becomes blue on large scales.

Lastly, we also consider the case where a super-inflationary 
phase is present initially $\tilde{p}<0$. In this case we do not need any 
particular assumption for the mass of the curvaton except for the condition
$\tilde{m}\ll \tilde{\cal H}$, and we also obtain a blue spectrum on large scales, 
as shown in Fig.~\ref{powerfinal}, though the blue tilt is not as sharp as
the case of the Jordan bounce.

\section{Conclusion}
\label{sec:conclusion}

To better understand the role of different conformal frames in cosmology,
we considered a simple analytical model in which a scalar field, an inflaton,
drives power-law inflation in the Einstein frame, and studied various Jordan 
frames associated with it by conformally transforming the metric,
where a Jordan frame is defined as the frame in which matter is minimally 
coupled to the metric while the inflaton has a non-minimal coupling with the
Ricci scalar. It should be noted that the predictions for both the tensor power 
spectrum and the curvature perturbation spectrum due to inflaton are 
unaltered in this setting. They are completely frame-independent
and the same as those for the standard power-law inflation.

Particular attention was paid to the physics from the matter point of view.
The minimal coupling of matter with a certain Jordan frame metric is
equivalent to a dilatonic coupling in the Einstein frame. But this simple
difference can lead to a completely different picture of the universe.
We studied two examples of how different the universe can be from 
the matter point of view. 

In section \ref{case1} we showed 
that matter can feel a super-inflationary universe or a 
decelerating contracting universe, even for a simple conformal 
transformation, inspired by the dilaton model \eqref{fcase1}, in spite of
 having inflation in the gravitational frame, i.e. the Einstein frame. 
Afterwards, as a representative of matter we considered
a curvaton which significantly contributes to the total scalar power spectrum, leaving 
imprints of its minimal coupling to a particular Jordan metric.
For instance, a blue tilt of the power spectrum on small scales
can be obtained if the curvaton feels super-inflationary expansion, 
as shown in Fig.~\ref{blue}, which can enhance the 
formation of primordial black holes.

In section \ref{case2}, we considered another particular conformal 
transformation \eqref{fcase2} which renders matter to feel a bouncing 
universe and therefore, as far as matter is concerned, the initial singularity 
is avoided. We obtained again the scalar power 
spectrum for a spectator curvaton, Fig.~\ref{powerfinal}, which is blue 
tilted on large scales. In the case where the curvaton mainly generates the 
total scalar power spectrum, this can give rise to an apparent suppression of
the power spectrum at large scales.

To conclude, we emphasise again that the purpose of this paper
is not to make specific predictions for a particular class of non-minimal 
coupling models. With this simple but analytic example what we want to 
stress is how much the matter point of view, i.e. the Jordan frame point of 
view, can differ from the gravitational point of view, i.e. the Einstein frame
point of view, and how this difference may actually affect observable
quantities like the curvature perturbation spectrum.

\begin{acknowledgments}
The authors would like to thank T. Tanaka for useful comments. 
This work was supported in part by the JSPS Grant in-Aid for 
Scientific Research (A) No.~21244033.
\end{acknowledgments}

\bibliography{bibliography}

\end{document}